\documentclass[fleqn,10pt]{wlscirep}

\usepackage[utf8]{inputenc}
\usepackage[T1]{fontenc}

\usepackage{bm}
\usepackage{mathrsfs}
\usepackage{amssymb}     
\usepackage{amsfonts}    
\usepackage{amsmath} 
\usepackage{greekctr} 
\usepackage{hyperref}
\usepackage{lmodern}

\title{Sub-8-nm resolution AKB-mirror-based hard X-ray ptychography via generalized Wirtinger projections}

\author[1,3,*]{Jie Dong}
\author[1,*]{Liang Zhou}
\author[1]{Zhongzhu Zhu}
\author[1]{Han Xu}
\author[1]{Aiyu Zhou}
\author[1]{Xuan Wang}
\author[1,2]{Shuo Wang}
\author[1,2]{Xiao Li}
\author[1,*]{Yuhui Dong}
\affil[1]{Institute of High Energy Physics, Chinese Academy of Sciences, Beijing 100049, China}
\affil[2]{University of Chinese Academy of Sciences, Chinese Academy of Sciences, Beijing 100049, China}
\affil[3]{Spallation Neutron Source Science Center, China Spallation Neutron Source, Guangdong 523803, China}

\affil[*]{e-mail: jiedong@ihep.ac.cn; zhouliang@ihep.ac.cn; and dongyh@ihep.ac.cn} 

\begin{abstract}
Owing to its unique capability for noninvasive, highly sensitive phase imaging of relatively large volumes at near-nanometre spatial resolution, hard X-ray ptychography has become increasingly essential in both the life and physical sciences. However, pushing resolution down to a few nanometres often requires highly customized, chromatic diffractive or refractive X-ray nanofocusing optics, significantly limiting the practical broadband energy-scan applications, such as spectro-ptychography. Here, we present the first known hard X-ray ptychographic imaging with a half-pitch resolution below 8 nm using total-reflection Advanced Kirkpatrick-Baez (AKB) mirror nanofocusing optics at the high energy photon source (HEPS), with clear potential for further extension. Despite leveraging the benefits of enhanced instrumentation, such as the high coherent flux of 4th-generation diffraction-limited storage rings (DLSR), state-of-the-art beamline X-ray optics and detectors, this is made possible by developing a reconstruction algorithm termed Generalized Wirtinger Projections (GWP). We derive the theory of GWP and experimentally demonstrate its capability for improved partial-coherence reconstruction and enhanced spatial resolution over conventional methods through imaging experiments on a Siemens star test chart at 12.4 keV ($\lambda \approx 1 \AA$). GWP provides a highly compact framework for jointly accounting for multiple coupled uncertainties that degrade resolution, enabling straightforward extension to other imaging modalities, such as burst ptychography, while delivering nearly an order-of-magnitude improvement in GPU memory efficiency. Furthermore, the ability to combine nanometre-scale spatial resolution with the inherently achromatic nature and high flux efficiency of AKB-mirror nanofocusing optics demonstrated in this work potentially opens new opportunities for \textit{in situ} or \textit{operando} broadband, energy-scan 3D spectroscopic imaging with element- or chemical-state specificity in complex environments at the nanoscale, holding significant promise for a wide range of applications from electronics and energy science to neuroscience.
\end{abstract}

\begin{document}

\flushbottom
\maketitle

\thispagestyle{empty}

\section*{Introduction}
X-rays are powerful for microscopy in that they offer high penetration, contrast specificity, and short wavelengths that potentially enable atomic-resolution imaging simultaneously. Equipped with powerful lenses such as ultrahigh-aspect-ratio hard X-ray zone plates, the latest transmission X-ray microscopy (TXM) techniques permit absorption-based imaging with a spatial resolution below 10 nm \cite{REF1, REF2}. However, TXM often requires additional optical components to obtain high-sensitivity phase-contrast imaging, which is indispensable at hard X-ray energies, particularly for low-atomic-number materials where phase contrast can exceed absorption contrast by orders of magnitude. Similar to coherent diffraction imaging (CDI) \cite{REF7}, ptychography circumvents the limitations of image-forming optics \cite{REF3,REF42,REF43} by leveraging the special properties of coherent wavefields, enabling lensless imaging with a resolution below 10 nm \cite{REF4, REF5, REF6}, while offering both absorption and quantitative phase contrast simultaneously. More specifically, ptychography operates by firstly encoding sample information in the Fraunhofer/Fresnel regions via interferometric/holographic approaches, next sequentially collecting the diffraction patterns from partially-overlapped illuminating probes to ensure sufficient phase diversity and data redundancy, then accurately modeling the forward physical process, such as light-matter interactions, light propagation or scattering, and lastly decoding the output by solving the phase problem with computational optimization routines. Many experimental constraints critical to X-ray microscopy can be relaxed in ptychography. Furthermore, additional experimental uncertainties, such as positioning errors \cite{REF8, REF9, REF10}, mixed states in both illumination and object \cite{REF11, REF12}, or multiple-scattering effects \cite{REF13}, can be addressed by exploiting the inherent data redundancy.

Pushing resolution down to a few nanometres with X-ray ptychography typically requires stringent experimental conditions across multiple aspects. Benefiting from stronger phase and absorption contrast as well as enhanced high-angle scattering, soft X-ray (200 eV-2 keV) ptychography enables imaging with a resolution approaching or even exceeding 5 nm \cite{REF14, REF15}. However, its applications are often limited to thin samples, due to the restricted penetration depth typically on the order of hundred nanometres \cite{REF15}. Hard X-rays (>5 keV) can penetrate samples up to $\sim$500 $\mu$m thick \cite{REF16}, yet achieving high-resolution ptychographic imaging at these energies remains challenging. Despite hardware stability on the nanometre scale, lenses with greater focusing power are generally required to access larger scattering angles, since X-rays are less strongly bent at higher energies. This constraint is usually satisfied by adopting diffractive or refractive optics, e.g., Fresnel zone plates (FZPs) with small outer zones \cite{REF6, REF17}, which naturally produce a wide angular diffraction cone, or multilayer Laue lenses (MLLs) with extremely small layer spacing \cite{REF4, REF18}, which provide even larger scattering angles than FZPs. Although characterized by a higher NA, these lenses are primarily chromatic and designed to operate over a narrow energy range, necessitating additional realignment for substantial energy shifts, while broadband energy scanning is increasingly indispensable for element- or chemical-specific imaging applications highly demanded across fields ranging from electronics to energy and neuroscience.

Kirkpatrick-Baez (KB) or more recent advanced Kirkpatrick-Baez (AKB) mirrors offer total-reflection nanofocusing of X-rays, and are typically characterized by no chromatic aberrations, high flux efficiency, long-term wavefront stability, and long working distances \cite{REF19, REF20, REF21, REF22}. However, the reflection geometry, particularly the grazing-incidence configuration, often imposes a fundamental limitation on the NA, leading to much weaker signals at high scattering angles and making high-resolution imaging challenging. A much lower signal-to-noise ratio (SNR) at high scattering angles can significantly degrade the final achievable image quality, pushing ptychography into a regime where noise and background become more dominant. Moreover, unlike FZPs or MLLs, AKB mirrors generally handle X-ray beams without spatial filtering, making a high degree of illumination coherence more critical. Therefore, achieving high resolution with a ptychography system using AKB mirrors typically requires X-ray sources with both a high degree of coherence and high brightness, resulting in its spatial resolution currently limited to the order of tens of nanometres \cite{REF21, REF22}. In contrast, more recent FZP- or MLL-based ptychography systems enable sub-10 nm hard X-ray imaging \cite{REF4, REF6}, and even a resolution approaching 4 nm in 3D has been reported for small volumes when combined with computed tomography (CT) \cite{REF6}.

Here, we report the first known hard X-ray ptychographic imaging with a resolution below 8 nm using total-reflection AKB nanofocusing mirrors at HEPS. This is made possible by leveraging the high coherent flux of the 4th-generation synchrotron source, advanced beamline arrangements, and specifically, by developing a ptychography reconstruction algorithm termed Generalized Wirtinger Projections (GWP). Characterized by a complex-valued, fully differentiable physical model and a high-performance, second-order phase-retrieval engine, GWP incorporates variable, task-specific loss functions to accommodate multiple experimental uncertainties that lead to significant resolution loss in ptychography, jointly accounting for photon-counting statistics, decoherence, and scan-positioning errors. We derive the principle and algorithm of GWP, experimentally demonstrate improved mixed-state reconstruction for partial-coherence X-ray illumination, and enhanced spatial resolution from hard X-ray ptychographic imaging results of a Siemens star test chart at 12.4 keV. Since it provides a highly compact framework that accounts for multiple coupled uncertainties simultaneously, GWP is widely compatible with other imaging modalities such as burst ptychography \cite{REF6}. Besides, GWP offers nearly an order-of-magnitude improvement in GPU memory efficiency compared to conventional methods. Furthermore, we believe the successful integration of high-resolution hard X-ray ptychography with achromatic nanofocusing optics demonstrated in this work potentially enables \textit{in situ} or \textit{operando} broadband, energy-scan 3D spectroscopic imaging with element or chemical-state specificity at the nanometre scale, facilitating wide-ranging applications spanning electronics, energy science, and neuroscience.

\section*{Results}

\begin{figure}[ht]
\centering
\includegraphics[width=\linewidth]{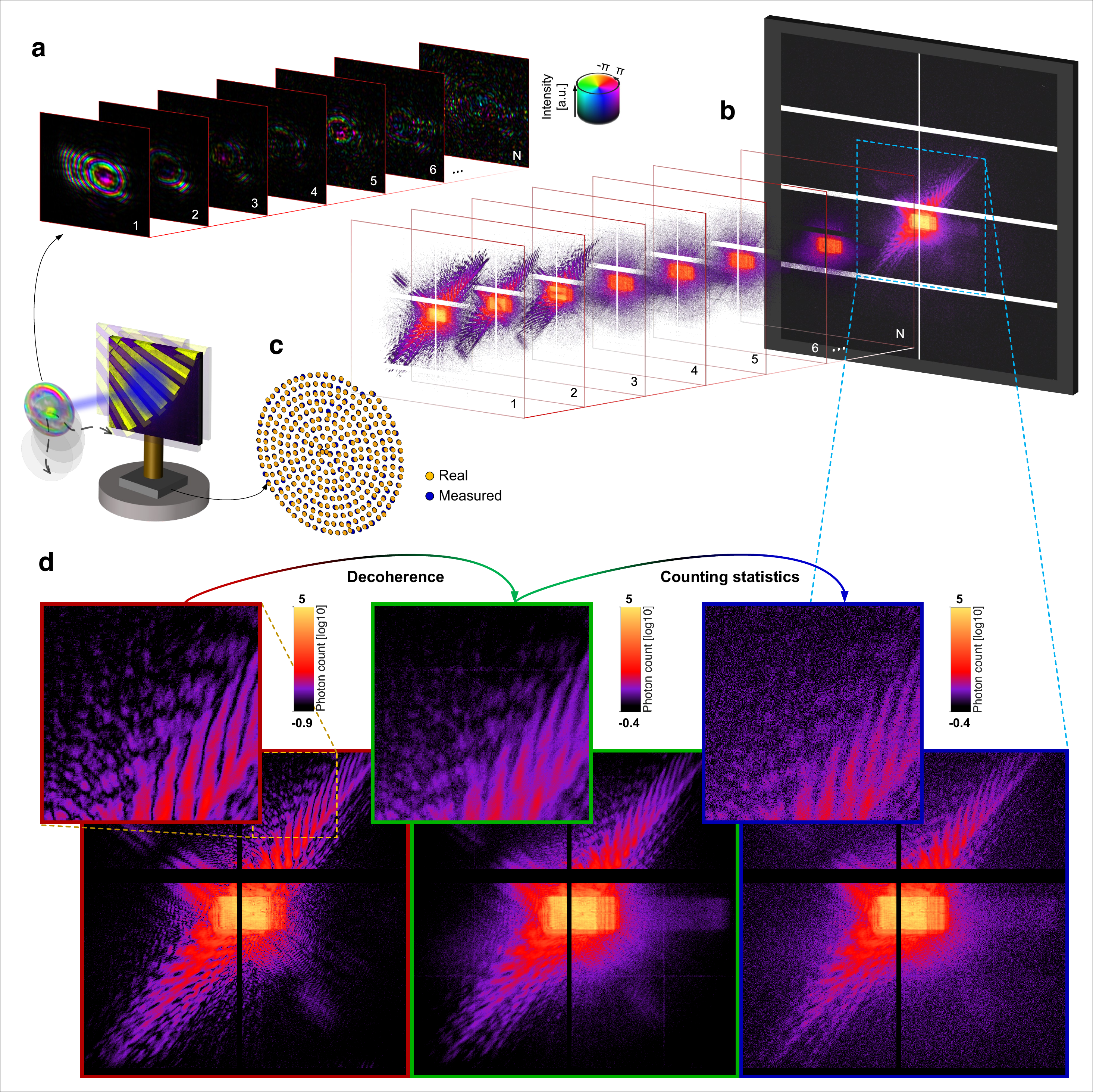}
\caption{\textbf{Illustration of the gap between digital forward models and physical reality that leads to imaging degradation in X-ray ptychography}. a, Schematic of the experimental setup for diffraction data acquisition, illustrating the physical layer of ptychography in which sample information is encoded in Fresnel/Fraunhofer diffraction. Decoherence arising from illumination instabilities, mixed states within the object, or detector fluctuations during a single exposure leads to state mixtures in both the probe and sample terms in real space, as well as in diffraction wavefields in reciprocal space. b, Poisson noise arises in the square-law detection of diffraction wavefields, given by the stochastic nature of photon counting. c, Positioning errors resulting from uncertainties in displacement tracking systems or probe drift during scanning. d, Compared with the ideal case of fully coherent illumination, diffraction patterns under partial coherence feature motion blur and a marked reduction in fringe (speckle) visibility. These effects are further exacerbated by counting noise, particularly in low-photon regions. The color bar indicates the number of detected photons on a logarithmic scale.}
\label{FIG1}
\end{figure}

\begin{figure}[ht]
\centering
\includegraphics[width=\linewidth]{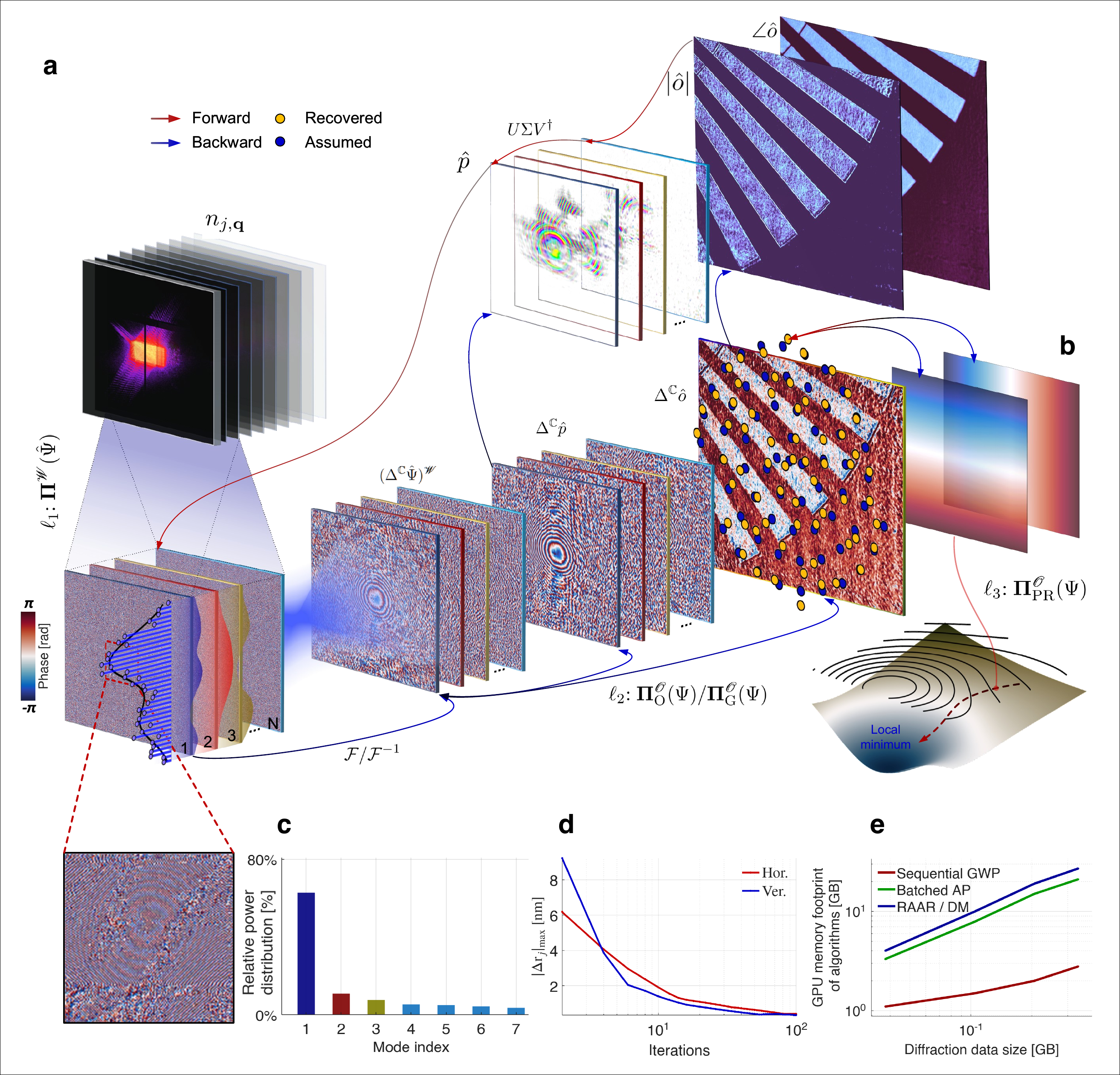}
\caption{\textbf{Principles, workflow, and computational efficiency of GWP for ptychographic reconstruction}. a, Complex-valued, fully differentiable physical model and high-performance, second-order phase-retrieval engine in GWP. Task-specific loss functions are incorporated, with explicit treatment of photon-counting statistics, mixed states, and relative positioning errors. The probe modes are orthonormalized by applying singular value decomposition (SVD) after each iteration or at defined intervals. b, Schematic of the embedded nonlinear optimization module for mixed-state scan position refinement. c, Schematic of the recovered relative power distribution of dominant coherence modes. d, Maximum absolute value of position correction versus iteration number, demonstrating convergence of the algorithm and successful nanoscale position refinement. e, Comparison of GPU memory footprints of the algorithms incorporating four mixed states, indicating nearly one-order-of-magnitude improvements in computational efficiency enabled by GWP over conventional algorithms. AP, alternating projections; RAAR, relaxed averaged alternating reflections.}
\label{FIG2}
\end{figure}
 
\begin{figure}[ht]
\centering
\includegraphics[width=0.9\linewidth]{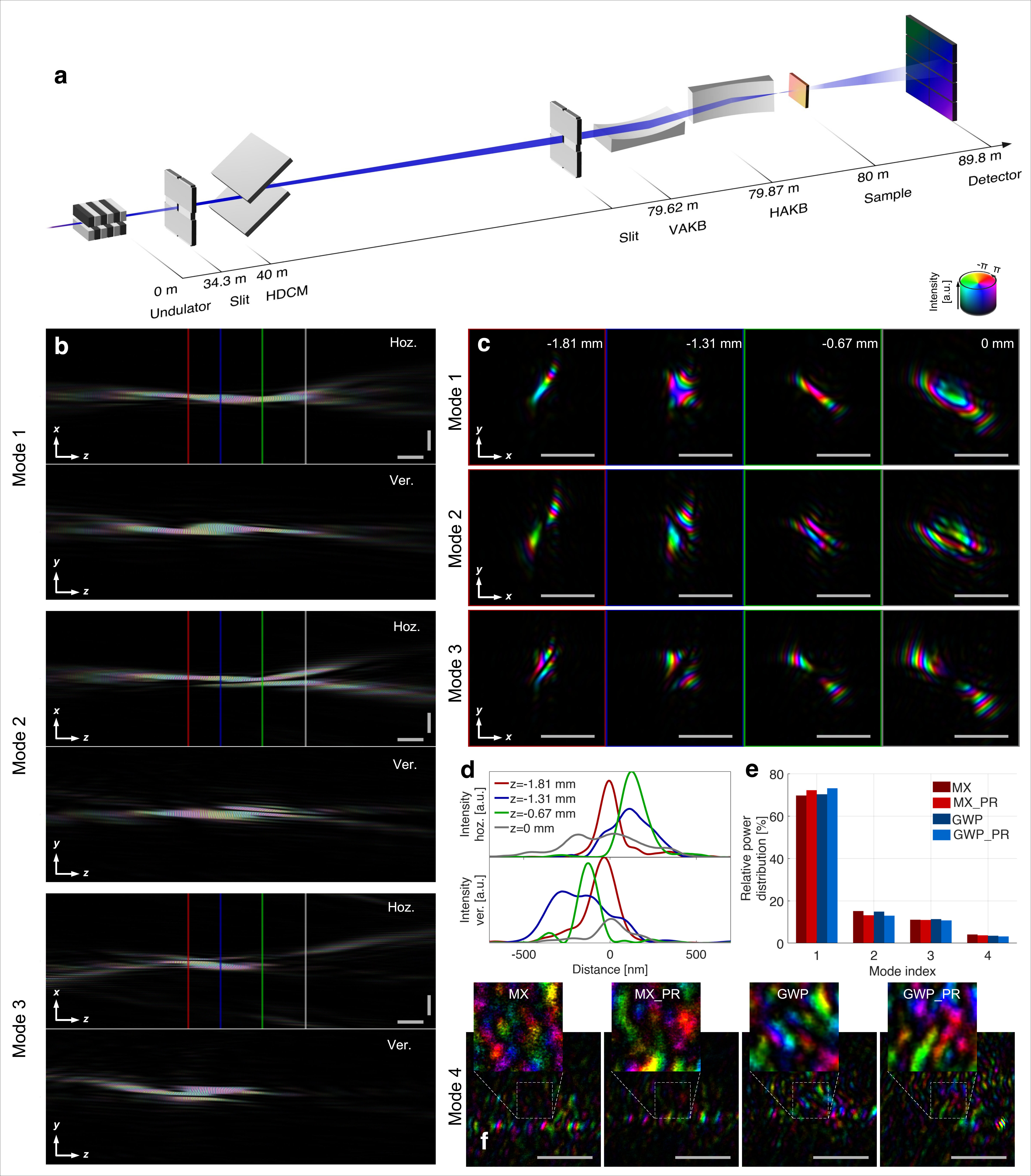}
\caption{\textbf{Mixed-state beam caustic characterization}. a, Schematic of the HXCS beamline with total-reflection AKB nanofocusing mirrors. b, 2D cross-sections of the beam wavefields near the focal region along the horizontal and vertical directions, corresponding to the first three dominant modes, respectively. The 3D beam wavefields are obtained via numerical propagation of mixed-state wavefields reconstructed using GWP. The red and green lines indicate the focal planes of the first and second AKB mirrors, respectively; the blue line corresponds to the intermediate plane; the gray line represents the sample plane. Scale bars are 0.4 mm and 1 $\mu$m in z and y direction, respectively. c, Wavefields at different focal planes along the propagation direction. Scale bar, 1 $\mu$m. d, The 1D cross-sections corresponding to mode 1 along the horizontal and vertical directions, respectively. e, Comparison of relative power distributions reconstructed using variable algorithms. MX, the conventional mixed-state reconstruction; PR, the reconstruction with the proposed position refinement module. The relative powers associated with the reconstructed first dominant modes are 69.69$\%$, 72.18$\%$, 70.25$\%$, and 73.11$\%$, respectively. f, Comparison of the fourth residual modes reconstructed using different algorithms. Scale bar, 1 $\mu$m. }
\label{FIG3}
\end{figure} 

\begin{figure}[ht]
\centering
\includegraphics[width=\linewidth]{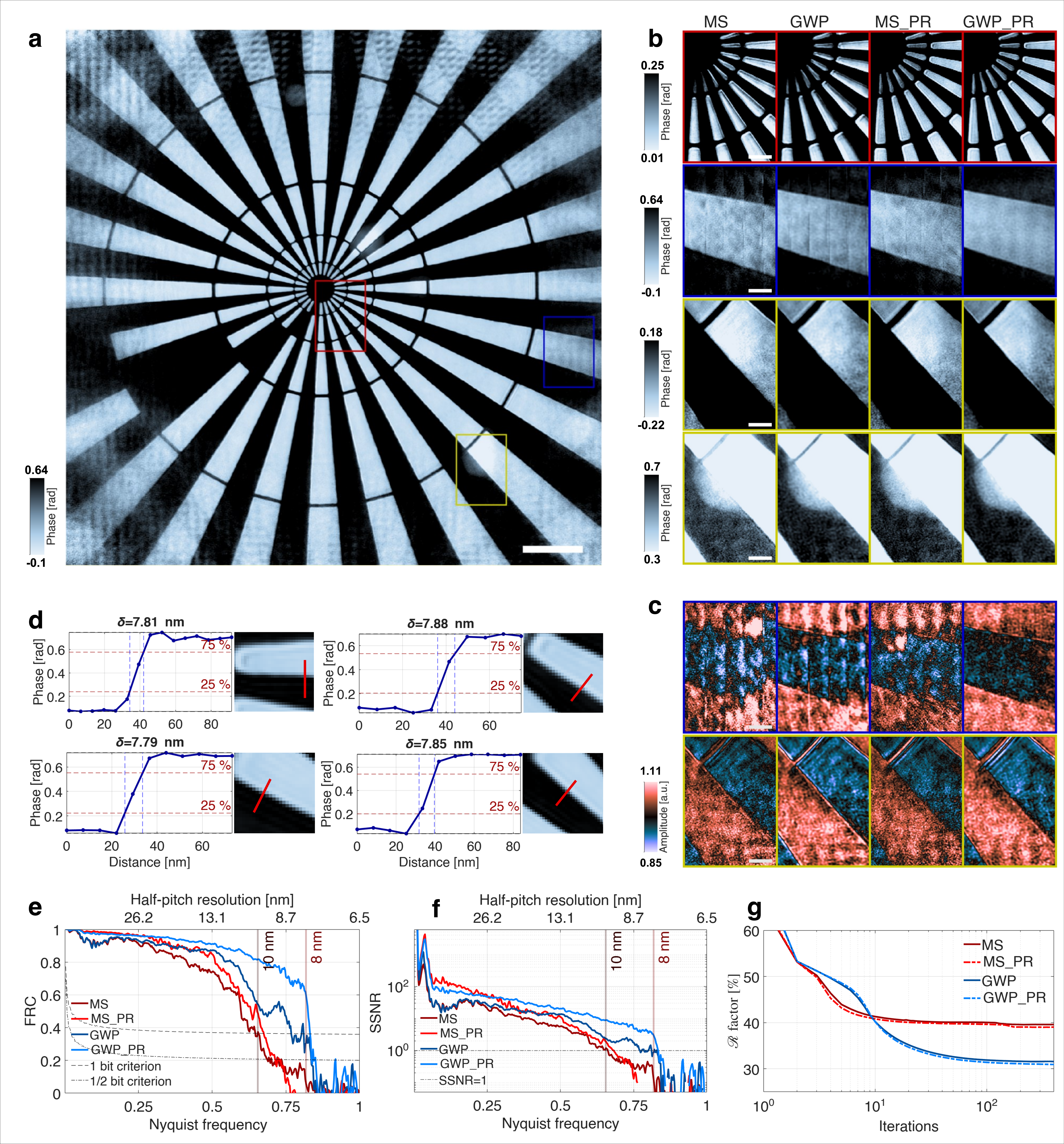}
\caption{\textbf{Sub-8 nm hard X-ray ptychographic imaging}.  a, Phase image from the GWP reconstruction of a Siemens star resolution target at the X-ray energy of 12.4 keV. Scale bar, 1 $\mu$m. b, Comparison of zoom-ins from boxed areas of the phase reconstructions obtained using different algorithms, but with the identical diffraction dataset and initial conditions. Columns 1 and 2 indicate the reconstructions without position refinement. Variable color scales are used to emphasize noise and reconstruction artifacts. Scale bar, 200 nm. c, Corresponding zoom-ins from the absorption reconstructions. Scale bar, 200 nm. d, Local-resolution estimations of the GWP reconstruction using edge profile analysis (EPA). e, Global-resolution estimations of all the reconstructions using Fourier ring correlation (FRC). f, Global-resolution estimations based on spectral signal-to-noise ratio (SSNR). g, R-factor versus iteration number over 400 iterations, indicating convergence behavior of the algorithms and demonstrating successful position refinement.}
\label{FIG4}
\end{figure}

\subsection*{Degradation model in X-ray ptychography}
Achieving exceptional spatial resolution with ptychography requires modeling the forward process as closely as possible to physical reality and accurately seeking a unique solution through algorithmic reconstruction (Fig. \ref{FIG1}). Most modern beamlines feature highly stable operating environments, high-precision alignment, interferometric scan positioning, and single-photon-counting data acquisition, yet unavoidable experimental factors from both environmental and instrumental sources, such as illumination variations, even at the tens-of-nanometres scale, can still blur diffraction patterns and significantly degrade reconstruction quality, especially when the imaging resolution approaches the nanometre scale.

\textbf{Decoherence}. The degree of coherence of the illuminating beam in AKB-mirror-based ptychography beamlines is strongly tied to the synchrotron source, thereby often leading to a coherence-limited spatial resolution \cite{REF11, REF12, REF23, REF24}. Although coherence can be improved by introducing filtering slits, the operation often reduces the beam flux by several orders of magnitude, thereby lowering the SNR. Moreover, achieving highly coherent measurements requires extremely stable optical components, ultrashort pulses, a static sample, and near-ideal detection, imposing stringent experimental constraints that are often difficult to satisfy simultaneously. Since it violates the assumption of fully coherent illumination in ptychography, decoherence can significantly reduce diffraction contrast (Figs. \ref{FIG1}a and \ref{FIG1}d) and further degrade reconstruction quality. Algorithmic approaches, such as mixed-state methods, have been shown to be effective and are therefore essential for mitigating decoherence during reconstruction.

\textbf{Counting statistics}. Single-photon-counting devices, such as EIGER detectors based on hybrid photon-counting technology, can effectively eliminate readout and dark noise, yet Poisson noise remains largely unavoidable in most X-ray scattering measurements \cite{REF25, REF26}. This factor is particularly critical for the ptychography system using AKB nanofocusing mirrors, where high-angle scattering is much weaker than in FZP- or MLL-based systems (Figs. \ref{FIG1}b and \ref{FIG1}d), necessitating accurate noise modeling. Inaccurate noise modeling obscures the definition of a unique solution in phase retrieval and directly leads to reduced convergence stability and resolution loss in ptychographic reconstruction \cite{REF27 }.

\textbf{Positioning errors}. Although the precision and stability of sample positioning during ptychographic scanning can be significantly improved using interferometric tracking systems of kHz-level frequencies, arising from instabilities in the monochromator, and more specifically from the nitrogen-cooled systems operating at several tens of Hz, probe drift on the order of tens of nanometres can still occur. The drift effect is effectively equivalent to scan-positioning errors and is normally difficult to measure directly, thereby potentially leading to a significant loss in resolution. 

\textbf{Multi-parameter coupling}. In practice, the factors mentioned above are often embedded within the same intensity measurements and are therefore strongly coupled through both the physical forward propagation and the digital reconstruction. For example, position correction relies on intensity gradients, which are often noise-limited, making the estimation ill-conditioned in low-dose regions. Besides, position errors and noise effects can also be absorbed into additional probe modes (see \hyperref[EXP]{ Experimental validation and characterization }). Hence, these different sources of uncertainty must be jointly accounted for.

\subsection*{Generalized Wirtinger Projections}
Mathematically, ptychographic phase retrieval can be formulated as an optimization problem, in which multiple parameters of interest are recovered simultaneously by iteratively searching for a unique set of predicted complex-valued exit waves $\hat{\psi}$ that minimizes the Euclidean distance $\ell$ between the reconstructed diffraction patterns $\hat{I}$ and the measured intensities $n$, while not violating additional constraints such as a fixed probe profile $\hat{p}$, or a continuous sample transmission function $\hat{o}$.
\begin{equation}\label{EQ1} 
\begin{aligned}
& \left( \hat{p}, \hat{o} \right) = \underset{\hat{p },\; \hat{o} \in  \mathbb{C}^{M \times N}} {\arg\min} \sum_j \sum_{\mathbf{q}} \sum_{l,k} \ell( \hat{I}_{j, \mathbf{q}},n_{j,\mathbf{q}}), \; \text{s.t.} \; \hat{I}_{j, \mathbf{q}} = \sum_{l,k}| \hat{\Psi}^{l,k}_{j,\mathbf{q}}  |^2 =  \sum_{l,k}|\mathcal{F}(\hat{p}^l_{j,\mathbf{r}} \hat{o}^k_{\mathbf{r}-\mathbf{r}_j} ) |^2,   
\end{aligned}
\end{equation}
where $\mathcal{F}$ denotes the Fourier transform , $j$ indexes the scan positions in ptychographic measurements, $\mathbf{q}$ and $\mathbf{r}$ denote the 2D coordinates in reciprocal and real space, respectively, and $l$ and $k$ index the individual probe and object coherence modes, accounting for possible state mixtures in both the illumination and sample term in X-ray scattering experiments.

Constraint sets are fundamental concepts to achieve successful phase retrieval in both CDI and ptychography \cite{REF28, REF29, REF30}. In most cases, they correspond  to the Fourier/modulus constraint in reciprocal space and the overlap constraint in real space, giving rise to two core subtasks in ptychographic optimization: searching for a unique set of $\phi$ that minimizes $\ell$; and decomposing $\phi$ into a common illumination profile and a spatially continuous sample transmission function, by exploiting the data redundancy encoded with ptychographic scanning.

To overcome the experimental imperfections that currently limit high-resolution ptychographic imaging with total-reflection nanofocusing optics, we implement GWP, a technique illustrated in Fig. \ref{FIG2}. Built on the constraint and projection framework, GWP is characterized by a complex-valued, fully differentiable forward model and a high-performance, second-order phase-retrieval engine that incorporates three task-specific loss functions to jointly account for photon-counting statistics, partial coherence, and positioning errors.

\subsubsection*{Solving $\ell$1: mixed-state Wirtinger projections}   
The mixed-state Fourier projection was first introduced by Thibault et al. \cite{REF11 }:
\begin{equation}   
\mathbf{\Pi}^{\mathscr{F}}(\Psi) :  \hat{\Psi} \rightarrow \hat{\Psi}^\mathscr{F} \triangleq { \frac{\sqrt{n_{j,\mathbf q}}} { \sqrt{\sum_{l,k}|{\hat{\Psi}^{l,k}_{j, \mathbf{q}}}|^2 } } }   {\hat{\Psi}^{l,k}_{j,\mathbf q} }.
\end{equation}  

The projection can be interpreted as a gradient-descent update based on an amplitude-based Gaussian log-likelihood loss \cite{REF31 }. However, in modern single-photon-counting detectors, the main noise during measurements is dominated by the Poisson distribution.  By replacing the conventional hard Fourier modulus projections with the Wirtinger projections we developed in our previous work \cite{REF31}, which incorporates noise-aware forward modeling and curvature-aware quantitative updating, GWP enables a direct, compact combination of accurate noise modeling and mixed-state reconstruction. More specifically, based on the Poisson probability distribution function (Eq. \ref{EQ3}a), $\ell$1 is defined as a real-valued negative log-likelihood function (Eq. \ref{EQ3}b): 
\begin{subequations} \label{EQ3} 
\begin{align}
P_{j,\mathbf{q}} (n \mid \hat{\Psi})             &= \frac{ \left(\sum_{l,k}|\hat{\Psi}^{l,k}_{j,\mathbf{q}}|^2  \right)^{n_{j, \mathbf{q}}}}{n_{j, \mathbf{q}}!} e^ { -\sum_{l,k}|\hat{\Psi}^{l,k}_{j,\mathbf{q}}|^2  },\\
\mathcal{L}(\hat{\Psi})  &=   -\log\prod_{j}\prod_{\mathbf{q}}  P_{j,\mathbf{q}} \triangleq  \sum_{j,\mathbf{q}}  \left( \sum_{l,k}|\hat{\Psi}^{l,k}_{j,\mathbf{q}}|^2 - {n_{j,\mathbf{q}}}\log{\sum_{l,k}|\hat{\Psi}^{l,k}_{j,\mathbf{q}}|^2}  \right).
\end{align}
\end{subequations} 
 
The mixed-state Wirtinger projection is then formulated to minimize $\mathcal{L}$ via nonlinear optimization:
\begin{equation} 
\mathbf{\Pi}^{\mathscr{W}}(\hat{\Psi}) :  \hat{\Psi} \rightarrow \hat{\Psi}^\mathscr{W} \triangleq  \underset{ \hat{\Psi} \in  \mathbb{C}^{M \times N} } {\arg \min} \mathcal{L}(\hat{\Psi}).
\end{equation}
  
Accordingly, the mixed-state WP updating can be defined as Eq. \ref{EQ5}b, by combining Newton’s method with the Hessian Wirtinger calculus and solving the linear equation, Eq. \ref{EQ5}a (see our previous work for a detailed derivation using CR calculus \cite{REF31 }):
\begin{subequations}  \label{EQ5} 
\begin{align}
(\Delta^\mathbb{C} \hat{\Psi})^\mathscr{W}   &= - \gamma ({ \mathcal{H}_\mathcal{L} ^\mathbb{C}  })^{-1}{ \nabla^\mathbb{C}\mathcal{L} } (\hat{\Psi}),\\
(\Delta^\mathbb{C}  \hat{\Psi}^{l,k}_{j,\mathbf{q}})^\mathscr{W}                  &= - \gamma  \frac{ {\sum_{l,k}|\hat{\Psi}^{l,k}_{j,\mathbf{q}}|^2 }  -  {n_{j,\mathbf{q}}}  }{ {\sum_{l,k}|\hat{\Psi}^{l,k}_{j,\mathbf{q}}|^2 } +  {n_{j,\mathbf{q}}}  } \cdot {\hat{\Psi}}^{l,k}_{j,\mathbf{q}},
\end{align}
\end{subequations}  
where $\nabla$ and $\mathcal{H}$ represent the Wirtinger gradient and Hessian of $\mathcal{L}$ with respect to $\hat{\Psi}$, respectively. $\gamma$ is a relaxation parameter.
 
\subsubsection*{Solving $\ell$2: parallel vs. sequential orthogonal overlap projections} 
The mixed-state decomposition can be formulated as the following minimization problem ($\ell$2), subject to the physical constraint of orthogonality among mixed states in partially coherent illumination:
\begin{equation} \label{EQ6}  
\mathbf{\Pi}^{\mathscr{O}}(\psi) :   \hat{\psi} \rightarrow \hat{\psi}^\mathscr{O} \triangleq \underset{\hat{p },\; \hat{o} \in  \mathbb{C}^{M \times N}}  {\arg \min} \left\|\hat{\psi}^{l,k}_{j,{\mathbf{r}}}-\hat{p}^{l}_{j,{\mathbf{r}}} \hat{o}^{k}_{\mathbf{r}-\mathbf{r}_j}\right\|_2^2, \;\text{s.t.} \;  \hat{p}^{m}_{j,{\mathbf{r}}} \cdot \hat{p}^{n}_{j,{\mathbf{r}}} = 0 \;(m\neq n),
\end{equation} 
where $\hat{\psi}$ denotes the inverse Fourier transform of $\hat{\Psi}$.

The problem can be solved using either a global or sequential strategy. Global approaches typically use the full set of diffraction patterns to perform batch updates at each iteration, incorporating all scan positions simultaneously. In contrast, sequential approaches update the probe and object using only a single diffraction pattern at a time, then they traverse all the scan positions iteratively in a randomized order. Analogous to the idea of the difference map (DM) method \cite{REF29, REF30}, a global solution can be obtained by setting the derivative of Eq. \ref{EQ6} with respect to $\hat{p}$ and $\hat{o}$ to zero:
\begin{subequations}   
\begin{align}
\mathbf{\Pi}_\mathrm{G}^{\mathscr{O}}(\Psi) : \hat{p}^l_{\mathbf{r}} &= \frac { \sum_j \{\sum_k\mathcal{F}[\hat{\Psi}^{l,k}_{j,\mathbf{q}}+ \eta (\Delta^\mathbb{C}  \hat{\Psi}^{l,k}_{j,\mathbf{q}})^\mathscr{W} ] \cdot {(\sum_k\hat{o}^k_{\mathbf{r}-\mathbf{r}_j})}^* \}} {\sum_j \left(|\sum_k\hat{o}^k_{\mathbf{r}-\mathbf{r}_j}|^2\right) }, \\
\hat{o}^k_{\mathbf{r}} &= \frac {\sum_{j} \{\sum_{l}\mathcal{F}[\hat{\Psi}^{l,k}_{j,\mathbf{q}} + \xi(\Delta^\mathbb{C}  \hat{\Psi}^{l,k}_{j,\mathbf{q}})^\mathscr{W}] \cdot {( \sum_{l}\hat{p}^l_{j,\mathbf{r}} )}^*\} } { \sum_{j} \left(|\sum_{l}\hat{p}^l_{j,\mathbf{r}}|^2\right) },
\end{align} 
\end{subequations}   
where $\eta$ and $\xi$ are the step sizes of the probe and object updating, respectively, and $*$ denotes complex conjugation.

The main drawback of global approaches for ptychographic reconstruction is that the memory footprint usually grows rapidly when mixed-state algorithms are incorporated. Although this issue can be partially mitigated by using mini-batch strategies, memory often becomes the primary limiting factor rather than computational time, particularly for GPU-based implementations. To address this challenge, an alternative sequential orthogonal overlap projection can be established by adopting an ePIE-style update scheme together with a regularization-based weighting strategy \cite{REF32}:
\begin{subequations}    
\begin{align}
\mathbf{\Pi}_\mathrm{S}^{\mathscr{O}}(\Psi) :\Delta \hat{p}^l_{j,\mathbf{r}} &= \eta \sum_k\frac { \mathcal{F}[(\Delta^\mathbb{C}  \hat{\Psi}^{l,k}_{j,\mathbf{q}})^\mathscr{W}] \cdot {(\hat{o}^k_{\mathbf{r}-\mathbf{r}_j})}^* } {(1-\alpha)|\hat{o}^k_{\mathbf{r}-\mathbf{r}_j}|^2 + \alpha|\hat{o}^k_{\mathbf{r}-\mathbf{r}_j}|_{\mathrm{max}}^2 },\\
\Delta \hat{o}^k_{\mathbf{r}-\mathbf{r}_j} &= \xi \sum_l\frac { \mathcal{F}[(\Delta^\mathbb{C}  \hat{\Psi}^{l,k}_{j,\mathbf{q}})^\mathscr{W}] \cdot {( \hat{p}^l_{j,\mathbf{r}} )}^* } {(1-\beta)|\hat{p}^l_{j,\mathbf{r}}|^2 + \beta|\hat{p}^l_{j,\mathbf{r}}|_{\mathrm{max}}^2 },
\end{align} 
\end{subequations}
where $\alpha$ and $\beta$ are the regularization parameters associated with $\hat{o}$ and $\hat{p}$, respectively.

The physical constraint of orthogonal probe states can be enforced by orthonormalizing $\hat{p}$ using singular value decomposition (SVD) (Fig. \ref{FIG2}), since partially coherent illumination can be modeled as a finite set of mutually independent fully coherent modes:
\begin{subequations} 
\begin{align} 
&\hat{p}_{j,\mathbf{r}}=\left[\hat{p}^1_{j,\mathbf{r}},\hat{p}^2_{j,\mathbf{r}},   \ldots, \hat{p}^N_{j,\mathbf{r}}\right] \in \mathbb{C}^{M \times N},\\
&\mathbf{\Gamma}(\hat p): \hat{p}_{j,\mathbf{r}} \rightarrow \hat{p}_{j,\mathbf{r}} ^\prime \triangleq U \Sigma V^\dagger = \left[{(\hat{p}^{1}_{j,\mathbf{r}})}^\prime,{(\hat{p}^{2}_{j,\mathbf{r}})}^\prime,   \ldots, {(\hat{p}^{N}_{j,\mathbf{r}})}^\prime\right]\in \mathbb{C}^{M \times N},
\end{align}
\end{subequations} 
where $U \Sigma V^\dagger$ represents the SVD of $\hat{p}$, and $\dagger$ denotes the conjugate transpose.

\subsubsection*{Solving $\ell$3: embedded position refinement for mixed states}
Numerical methods have been shown to be effective for correcting positioning errors in ptychographic reconstruction, and subpixel accuracy has already been demonstrated under fully coherent illumination \cite{REF8, REF9, REF10, REF33}. Here, we propose an embedded position refinement method based on the Gauss-Newton algorithm, specifically designed for mixed-state reconstruction, thereby offering both high memory efficiency and broad applicability.

We formulate the position refinement task as the following minimization problem $\ell$3 by utilizing the real-space overlap constraints: 
\begin{equation}  
\mathbf{\Pi}_\mathrm{PR}^{\mathscr{O}}(\Psi) : \left( \hat{\mathbf r}^x_j, \hat{\mathbf r}^y_j \right) \triangleq \underset{\hat{\mathbf r}_j \in  \mathbb{R}} {\arg \min} \sum_{\mathbf{q}}\sum_{l,k}  \mathcal{R}_{j,\mathbf{q}}(\hat{\Psi}^{l,k}_{j,\mathbf{q}},n_{j,\mathbf{q}}), \; \text{s.t.} \; \hat{\Psi}^{l,k}_{j,\mathbf{q}} = \mathcal{F}(\hat{p}^{l}_{j,\mathbf{r}}\hat{o}^{k}_{\mathbf{r}-\hat{\mathbf r}^{l,k}_j}),
\end{equation}    
where $\mathcal{R}$ denotes the residual term, which is defined as the following sum of squared errors:
\begin{equation}  \label{EQ11} 
{\mathcal{R}}_{j,\mathbf{q}}  =  (\sum_{l,k}|\hat{\Psi}^{l,k}_{j,\mathbf{q}}|^2-n_{j,\mathbf{q}})^2.
\end{equation}    

The Jacobian matrix of Eq. \ref{EQ11}, corresponding to the first derivative of $\mathcal{R}$ with respect to $\hat{\mathbf r}$, can be formulated as follows by using the Fourier shift theorem:
\begin{equation}    
{\mathcal J}^{l,k}_{j,\mathbf{q}}  = \frac{\partial \mathcal{R}_{j,\mathbf{q}}}{\partial{\hat{\mathbf r}_{j,\mathbf{q}}} } =    4( \sum_{l,k}|\hat{\Psi}^{l,k}_{j,\mathbf{q}}|^2-n_{j,\mathbf{q}}) \Re \left[ ({\hat{\Psi}^{l,k}_{j,\mathbf{q}}})^* \mathcal{F}\left\{ \hat{p}^l_{j,\mathbf{r}} \cdot {\mathcal{F}^{-1} [-i2\pi {(N_x+N_y)}\cdot\mathcal{F} (\hat{o}^k_{\mathbf{r}}) e^{-i2\pi{(N_x+N_y)}\hat{\mathbf r}^{l,k}_j} ] }\right\}\right],
\end{equation}    
where $N_x$ and $N_y$ are linear ramp vectors associated with the horizontal and vertical directions, respectively.

Then, the quantitative update step can be extracted by solving the least-squares problem using the Gauss-Newton algorithm:
\begin{equation}  
\Delta\hat{\mathbf r}_j = -\zeta \frac{\sum_{\mathbf{q}}{\mathcal J}^{l,k}_{j,\mathbf{q}} {\mathcal{R}}_{j,\mathbf{q}}  }{\sum_{\mathbf{q}} |{\mathcal J}^{l,k}_{j,\mathbf{q}}|^2 },
\end{equation}     
where $\zeta$ denotes the relaxation parameter for the position updates.

\subsection*{Experimental validation and characterization} 
\phantomsection 
\label{EXP}
To validate the performance of GWP on experimental datasets, we conducted hard X-ray ptychographic imaging experiments on a Siemens star resolution test chart, which was fabricated by forming $\sim$620 nm-thick gold (Au) absorber patterns on silicon nitride (SiN) membranes and featured a minimum half-pitch size down to 30 nm (see \hyperref[METHODS]{Methods} for details). The tests were performed at the HXCS beamline of HEPS \cite{REF41} (see \hyperref[METHODS]{Methods} for details). As shown in the optical layout (Fig. \ref{FIG3}a), the beamline was characterized by an undulator source and a pair of AKB mirrors positioned at a distance of 79.6 m from the source for focusing the beam to extremely small, highly coherent spots, typically down to a few micrometers.

Two groups of datasets, including a total of 961 $\times$ 2 diffraction patterns, were collected using a slightly defocused X-ray probe of $\sim$1 $\mu$m in diameter at 12.4 keV ($\lambda \approx 1 \AA$), with an exposure time of 1s (see \hyperref[METHODS]{Methods}). The ptychographic scanning followed a snake-like raster pattern with a step size of $\sim$150 nm. Initialized with four probe modes, variable algorithms for ptychographic reconstruction were applied to the same diffraction datasets (see \hyperref[METHODS]{Methods}).

\subsubsection*{Mixed-state focus characterization} 
To validate the performance of GWP in handling partially-coherent illumination, we first performed mixed-state focus characterization based on a single diffraction dataset, considering ptychography is one of the most important methods for characterizing nanofocused X-ray beams and quantifying aberrations in X-ray optics \cite{REF34, REF35}. The 3D beam caustic (Fig. \ref{FIG3}b) was reconstructed by retrieving multiple coherence modes of the illumination first and then subsequently extending the retrieved wavefields backward and forward in free space along the propagation direction using the angular spectrum method (see \hyperref[METHODS]{Methods}).

The beam caustic results indicated that the line-shaped focused spots associated with the two AKB mirrors were at 1.8 mm and 0.67 mm upstream, respectively. The two spots exhibited a full width at half maximum of $\sim$150 nm, as illustrated by the recovered complex-valued wavefields (Fig. \ref{FIG3}c) and the corresponding 1D cross-sectional profiles (Fig. \ref{FIG3}d). There was a square-shaped focal spot in the middle, while the illuminating probe recovered at the object plane was slightly out of focus. The retrieved probe included four modes (Figs. \ref{FIG3}c, \ref{FIG3}e, and \ref{FIG3}f), in which the first mode exhibited a dominant relative power of around 70$\%$, while the second and third modes were characterized by low-intensity orthogonal sidelobes along the optical axis, indicating relatively stable illumination provided by the AKB-mirror nanofocusing system during exposure. For comparison, we further repeated the caustic retrieval process with various reconstruction algorithms. The results indicated similar distributions of the dominant coherence modes (Fig. \ref{FIG3}e), validating the successful mixed-state ptychographic reconstruction of GWP.

Additionally, in both the MX and GWP reconstructions, the degree of illumination coherence increased after applying the proposed position refinement algorithm, suggesting that the reconstruction algorithms previously interpreted positional errors as spatial variations and absorbed them into the coherence modes. Furthermore, for both reconstructions, with and without position refinement, GWP showed higher coherence compared to the conventional MX method. Based on a mechanism similar to that in position refinement, it can be inferred that the conventional algorithm treated the noise effects in the diffraction patterns as additional decoherence sources and absorbed them into extra modes. This effect was also clearly observed in the complex profile comparison of the residual modes (Fig. \ref{FIG3}f), where conventional reconstructions exhibited increased speckle visibility, whereas the GWP results were noticeably smoother and featured lower relative powers.

Hence, it can be concluded that by replacing hard projections with soft, statistical projections, GWP weights weak-scattering signals more appropriately and reshapes the landscape of searching space, thereby improving the decoupling of noise and other ambiguities and enabling a more accurate characterization of partial coherence.

\subsubsection*{Sub-8-nm resolution Hard X-ray imaging by GWP} 
We further validated the improved resolution of GWP by comparing hard X-ray ptychographic imaging results from the resolution target, using both qualitative and quantitative evaluation approaches. The reconstructed phase image, corresponding to the projected electron density along the X-ray propagation direction after linear phase ramp removal and phase unwrapping, spanned a field of view of $\sim$11 $\mu$m $\times$ 11 $\mu$m (Fig. \ref{FIG4}a) and encompassed the finest features of the sample at the center for reliable resolution analysis.

Following the same comparing strategy as in the focus characterization analysis, significant improvements in image quality after applying the position refinement module (columns 3 and 4) were clearly observed in the zoom-ins (Fig. \ref{FIG4}b) from the regions marked with red and blue boxes (Fig. \ref{FIG4}a), thereby further confirming the successful position refinement.

We qualitatively evaluated the imaging performance of GWP by comparing reconstructions obtained with various algorithms. Significant noise suppression was observed in the zoom-ins marked by the red box, corresponding to the central area of the reconstruction, where data redundancy was close to its maximum. Clear suppression of artifacts was evident in the zoom-ins marked by the blue box, located near the edges of the field of view, where data redundancy was reduced. Moreover, noise suppression across different materials was demonstrated in the zoom-ins marked by the yellow box. The region contained a gold droplet of $\sim$500 nm in diameter formed randomly by fabrication imperfections, and thus exhibited richer phase variations. Variable color scale ranges were applied to highlight subtle phase fluctuations, and consistent noise suppression in both the patterned regions and the background, corresponding to the gold absorber and silicon nitride (SiN) membrane, respectively, was clearly revealed (Fig. \ref{FIG4}b). More significant differences were evident in the absorption reconstructions (Fig. \ref{FIG3}c), given by the sample’s intrinsic dynamic range spanning more than one order of magnitude between $\beta$ and $\delta$. Although the absorption images were of lower overall quality than the phase reconstructions, the GWP results still demonstrated noticeable improvements in both high-frequency noise suppression and artifact reduction. These improvements were particularly evident in regions of low data redundancy and inherently limited SNR (row 1 in Fig. \ref{FIG4}c).

We quantitatively evaluated the imaging performance of GWP using both global and local approaches. Since ground truth was unavailable, global-resolution analysis was performed by processing two independent ptychographic datasets and calculating the FRC \cite{REF36} and SSNR \cite{REF37, REF38} curves (Figs. \ref{FIG4}e and \ref{FIG4}f). In the FRC comparison, the results of GWP demonstrated an approximate 1.24$\times$ improvement in resolution over the MX reconstructions, improving from $\sim$9.7 nm to $\sim$7.8 nm based on the 1-bit criterion. Consistently, the SSNR analysis indicated a substantial gain approaching one order of magnitude in the high-resolution range, corresponding to spatial frequencies from 1/10 to 1/8 nm$^{-1}$. Furthermore, since the reconstructed images contained high-contrast, well-defined features with sharp edges, we further performed local-resolution evaluations using EPA \cite{REF39, REF40} based on the 25-75$\%$ edge response criterion, which is equivalent to half the Rayleigh resolution. The EPA results directly confirmed sub-8 nm half-pitch resolution in real space (Fig. \ref{FIG4}d), in good agreement with the global FRC results. In addition, the successful position refinement and the improved performance of GWP were reflected in the R-factor curves, which characterized the overall agreement between the reconstructed diffraction intensities and the measured values as a function of iteration number. Although not a direct measure of resolution, these curves clearly indicated that GWP reconstructions achieved a better overall fit to the measured data (Fig. \ref{FIG3}g).

\section*{Discussion}

We successfully demonstrate sub-8-nm resolution hard X-ray ptychographic imaging using total-reflection, achromatic AKB mirrors nanofocusing optics at the HXCS beamline of HEPS. Despite leveraging the advances in 4th-generation synchrotron sources, X-ray optics, and detectors, this was enabled by GWP, a high-performance reconstruction algorithm that we developed. GWP offers a compact framework for directly combining three key components crucial to achieving high-resolution hard X-ray ptychography with AKB mirrors nanofocusing optics: the Wirtinger projections for accurate noise modeling, the mixed-state algorithm for handling partially coherent illumination, and the embedded position refinement module to correct relative probe-sample drift during scanning. We present the detailed GWP algorithm, along with comparisons of reconstruction results obtained with variable algorithms. In the experiments, we have demonstrated improved mixed-state reconstruction and enhanced spatial resolution down to 8 nm enabled by GWP from hard X-ray ptychographic imaging on a Siemens star test chart at 12.4 keV. GWP offers nearly an order-of-magnitude improvement in GPU memory efficiency and allows straightforward extension to other imaging modalities, such as burst ptychography. Furthermore, the successful integration of high-resolution ptychography with achromatic AKB mirrors nanofocusing optics demonstrated in this work potentially enables \textit{in situ} or \textit{operando} nanometre-scale, energy-scan spectroscopic imaging with element or chemical-state specificity across a broad energy range, holding significant promise for wide applications ranging from electronics and energy science to neuroscience.

We believe the current results clearly indicate substantial potential for further improvements in spatial resolution, since not all instrumental and experimental parameters have been pushed to their extreme limits. To achieve this goal, it’s straightforward to decrease the sample-to-detector distance, thereby extending the coverage of higher-scattering angles in reciprocal space and reducing the reconstructed pixel size in real space. Although this operation also reduces the oversampling ratio, the issue can be mitigated by further reducing the probe size to ensure that the far-field diffraction patterns are adequately sampled by the detector, thereby preserving reconstruction quality. Additionally, given that poor SNR at high scattering angles is often a limiting factor in the current system, directly increasing the number of incident photons, e.g., increasing the X-ray coherent flux, appears helpful. Experimentally, this method should be accessible, since the current peak count rate is approximately 1$\times 10^6$ photons/pixel/s, much lower than the maximum value allowed by the present detector, $10^7$ phs/s/pixel. Another possible approach is to increase the exposure time; however, the trade-off between high-angle SNR and the degree of illumination coherence must be carefully considered, as coherence can degrade significantly due to accumulated vibrations over longer exposures. Additionally, it should be noted that the detector dynamic range, rather than the coherent flux, is more likely to become a primary limiting factor for achieving higher resolution with ptychography, since X-rays with high brightness become increasingly available at modern high-brilliance synchrotron and free-electron laser (XFEL) sources. We believe this problem can be addressed by introducing customized beamstops, such as semi-transparent ones, to partially or totally attenuate the direct beam. Together with the further development of corresponding algorithms, this approach may enable more efficient use of the detector’s dynamic range, therefore permitting further gains in the achievable imaging resolution.

\section*{Methods}
\phantomsection 
\label{METHODS}
\subsection*{Experimental set-up and data acquisition}
The experiment was conducted at the HXCS beamline of HEPS. As a member of 4th-generation DLSRs, HEPS features a circumference of 1,360 m, a maximum electron energy of 6 GeV, and an emittance as low as 60 pm, which is particularly important for methods requiring high coherence, such as ptychography \cite{REF41}. The facility delivers X-rays with a photon energy of up to 300 keV and a peak brightness of up to 4 $\times$ $10^{21}$ phs/s/mm$^2$/mrad$^2$/0.1$\%$ BW. 

The HXCS beamline is designed to take full advantage of the high brilliance and coherent flux of HEPS and mainly focuses on coherent diffraction imaging with nanometre-level spatial resolution based on total-reflection AKB mirrors nanofocusing optics (Fig. \ref{FIG3}a). The beamline operates over the energy range 7-25 keV. X-rays generated by an undulator source are sequentially shaped by a slit located $\sim$34.3 m downstream and filtered by a double-crystal monochromator (DCM) positioned $\sim$40 m downstream. The DCM can use either a Si(111) double-crystal configuration for tunable energy operation or a Si(311) channel-cut configuration for enhanced stability. After passing through the DCM, the beam is further filtered by a slit and focused by a pair of AKB mirrors positioned $\sim$79.62 m and $\sim$79.84 m downstream for vertical and horizontal focusing, respectively.

The specimen used in our experiments was a Siemens star test chart (Applied Nanotools Inc., ANT), in which $\sim$620 nm-thick gold (Au) absorber patterns were fabricated on $\sim$1 $\mu$m-thick silicon nitride (SiN) membranes, with a minimum half-pitch of 30 nm. The specimen was mounted on a piezoelectric stage with a positioning precision of $\sim$2 nm. 

The diffraction patterns were recorded using an EIGER2 XE 4M detector (Dectris Ltd.), a hybrid photon-counting X-ray detector with negligible readout noise and dark current. The detector had 2068 $\times$ 2162 pixels with a pixel size of 75 $\mu$m, and a maximum count rate of $10^7$ ph/s/pixel. It was positioned 9.8 m downstream of the specimen, with an evacuated flight tube installed between the sample and detector. We used X-rays at 12.4 keV for illumination, monochromatized by a channel-cut Si(311) crystal. The illuminating probe was slightly defocused at the sample plane, resulting in a $\sim$1 $\mu$m-diameter spot. Two diffraction datasets, each consisting of 961 diffraction patterns, were collected. The scan followed a snake-like raster trajectory with a step size of $\sim$150 nm. To mitigate the raster-grid pathology, random position offsets with a standard deviation of 10$\%$ ($\sim$15 nm) were introduced. For each scan position, the exposure time was 1 s with an additional dead time of 0.99 s (1.99 s total acquisition time), and the maximum photon count was $\sim$1.3 $\times$ $10^6$ ph/s/pix, resulting in a total of $\sim$ 3.7 $\times$ $10^8$ photons per diffraction pattern. The total acquisition time for a single diffraction dataset was approximately 0.5 h.

\subsection*{Data reconstruction}
All reconstructions presented in this work were performed with a self-developed GPU-accelerated package for ptychographic phase retrieval. Arrays of 1996 $\times$ 1996 pixels were subsequently cropped from the original diffraction patterns and binned by 2 $\times$ 2, resulting in a pixel size of $\sim$6.5 nm in the reconstructed object plane. To verify the generality, the object was initialized with a uniform intensity distribution and a random phase distribution. The initial probe was modeled as a flat circular with a Gaussian phase profile, with the caustic adjusted to match typical experimental conditions. Four probe modes were applied to account for partial coherence effects in illumination. \textit{Ab initio} phase retrieval, i.e., joint optimization accounting for both noise and decoherence effects, was performed by using GWP, rather than through conventional ptychographic reconstruction sequentially followed by maximum-likelihood refinement \cite{REF27}.

All computational procedures were implemented in MATLAB R2025b with NVIDIA CUDA v12.0. The platform was deployed on a desktop workstation equipped with a single NVIDIA GeForce RTX 5090 GPU (32 GB VRAM), an Intel Core Ultra 9 285K processor, and 128 GB of system memory. For the sequential GWP reconstruction with a single-precision floating-point number (32 bits, 4 bytes), the typical time for processing a diffraction dataset of size 841 $\times$ 970 $\times$ 970 ($\sim$2.9 GB) on the single RTX 5090 GPU was $\sim$1.7 s per iteration, with a GPU memory footprint of $\sim$18.5 GB.

\subsection*{Focus characterization}
The complex-valued wavefields associated with mixed states in illumination were reconstructed using both the GWP and MX algorithms. Thereafter, by using the angular spectrum method \cite{REF39} derived from the scalar Helmholtz equation, the complex wavefields corresponding to each coherence mode can be calculated separately at arbitrary positions along the optical axis, thereby revealing the beam caustic:
\begin{equation}    
\hat{p}^l_{j,(\mathbf{r},z)}=\mathcal{F}^{-1}\left[\mathcal{F}\left(\hat{p}^l_{j,\mathbf{r}}\right)  e^{i k_z z}\right], 
\text{ } k_z={2 \pi}\sqrt{\left({1}/{\lambda}\right)^2-f_x^2-f_y^2},
\end{equation} 
where $f_x$ and $f_y$ denote the spatial frequency coordinates, and $\lambda$ represents the wavelength of illuminating X-rays.

\subsection*{Resolution estimation}
FRC provides quantitative estimates of global spatial resolution for various imaging systems and is increasingly regarded as the gold standard across multiple disciplines. By calculating correlations over corresponding rings (2D) or shells (3D) in reciprocal space, FRC quantifies the similarity between two independent reconstructions as a function of spatial frequencies. The 1/2-bit and more stringent 1-bit threshold are widely used resolution criteria in FRC analysis, indicating the highest spatial frequency at which the corresponding amount of reliable information is still retained. In this work, we quantitatively evaluated the performance of different algorithms by calculating the FRC curves between complex-valued object transmission functions reconstructed from two independent diffraction datasets acquired within several hours under nearly identical experimental conditions. Prior to FRC analysis, real-space registration and phase-ramp removal were carefully performed to ensure accurate alignment between the two images.

A more detailed quantitative assessment of reconstruction quality can be further obtained by calculating the SSNR curves \cite{REF37, REF38} based on the FRC results:
\begin{equation}    
\operatorname{SSNR}(\mathbf{q})= \frac{2\operatorname{FRC}(\mathbf{q})}{1-\operatorname{FRC}(\mathbf{q})}.
\end{equation}  

EPA is a straightforward approach that estimates local spatial resolution in real space by measuring the sharpness of edges that appear in an image. The resolution is typically determined by the 25-75$\%$ criterion \cite{REF39, REF40} in EPA. Although less sensitive to background offsets and phase misalignment, EPA additionally requires good edge quality of the features for analysis.
\\

\noindent\textbf{Acknowledgements}\\ 
This work was funded by the Beijing Natural Science Foundation (Grant No. 1244063), the National Natural Science Foundation of China (Grant No. 12405372), and the China Postdoctoral Science Foundation (Grant No. 2024M751758). The authors also gratefully acknowledge support from the High Energy Photon Source (HEPS) project. \\ \\

\noindent\textbf{Author contributions}\\
J.D. proposed the concept, developed the underlying principles and algorithms, conducted the experiments, analyzed the data, and contributed to writing the manuscript. L.Z., Z.Z., H.X., and X.W. designed the experimental setup and conducted the experiments. S.W. and X.L. also contributed to the experiments. Y.D. supervised the project. All authors reviewed the manuscript. \\

\noindent\textbf{Competing interests}\\
The authors declare no competing interests. \\

\noindent\textbf{Data availability}\\
The data that support the findings of this research are available from the corresponding authors upon reasonable request.\\

\nocite{*} 
\bibliography{n8}

\end{document}